\documentclass[conference]{IEEEtran}
\IEEEoverridecommandlockouts
\usepackage{cite}
\usepackage{amsmath,amssymb,amsfonts}
\usepackage{algorithmic}
\usepackage{graphicx}
\usepackage{textcomp}
\usepackage{xcolor}

\usepackage{subfig}

\graphicspath{ {images/} }

\def\BibTeX{{\rm B\kern-.05em{\sc i\kern-.025em b}\kern-.08em
    T\kern-.1667em\lower.7ex\hbox{E}\kern-.125emX}}
\begin{document}

\title{End-to-End Optimization of JPEG-Based Deep
Learning Process for Image Classification
\thanks{This material is based upon work supported by Google Cloud.}
\vspace*{-2mm}

}

\author{\IEEEauthorblockN{Siyu~Qi,
    Lahiru~D.~Chamain,
	and~Zhi~Ding}
\IEEEauthorblockA{Department of Electrical and Computer Engineering\\ University of California, Davis, CA, 95616\\Email: \{syqi, hdchamain, zding\}@ucdavis.edu.
\vspace*{-2mm}
}
}

\maketitle

\begin{abstract}
Among major deep learning (DL) applications, distributed learning involving image classification require effective image compression codecs deployed on low-cost sensing devices for efficient transmission and storage. Traditional codecs such as JPEG designed for perceptual quality are not configured for DL tasks. This work introduces an integrative end-to-end trainable model for image compression and classification consisting of a JPEG image codec and a DL-based classifier. We demonstrate how this model can optimize the widely deployed JPEG codec settings to improve classification accuracy in consideration of bandwidth constraint. Our tests on CIFAR-100 and ImageNet also demonstrate improved validation accuracy over preset JPEG configuration.
\end{abstract}


\vspace*{-2mm}

\section{Introduction}
In recent years, deep convolutional neural networks (CNNs) have demonstrated successes in learning
tasks such as image classification and recognition, 
owing to their capability of extracting image features among adjacent pixels. The emergence of residual network (ResNet)\cite{He_2016} further enhanced 
image classification without introducing extra 
computational complexity.

In the era of IoT and cloud computing, many practical applications rely on widely deployed low-cost cameras and
sensors for data collection before 
transmitting sensor data to powerful cloud or edge 
servers that host pre-trained deep classifiers. 
As most (RF) network links usually are severely
band-limited and must prioritize heavy data traffic, image compression 
techniques are vital for efficient and effective utilization of limited
network bandwidth and storage resources. 
JPEG~\cite{itut81} is a highly popular codec 
standard for lossy image compression, widely used to conserve bandwidth in source data transmission and storage. 
The JPEG encoding process includes discrete cosine transform (DCT) and 
quantization. 
The quantized integer DCT coefficients are encoded via
run-length encoding (RLE) and Huffman coding. Due to RLE, total bit rate of an image cannot be predicted straightforwardly\cite{wu1993rate,tuba2017jpeg,luo2020rate}.
The JPEG encoding achieves substantial image compression ratio with little human perception quality sacrifice. These encoded bits are then transmitted over a channel of limited capacity (bit rate) before decoding and recovery for various applications.

Targeting human users, the parameters in JPEG configuration
are selected according to visualization subjective tests. However, in CNN-based image classifications, na\"ive adoption of the lossy JPEG image encoding, designed primarily for
human visualization needs, can lead to unexpected accuracy loss because the traditional CNN models are agnostic of the compression distortion. To tackle this issue, this work is motivated by the obvious and important question in
distributed AI: \textit{How to optimally (re)configure standardized JPEG for image compression
to improve DL-based image classification.} 

Motivated by the strong need to conserve 
network bandwidth and local storage for remote image classification, we present an end-to-end trainable 
DL model for joint image compression and classification that can optimize the widely deployed JPEG codec to improve classification accuracy over current JPEG settings. 
We formulate this dual-objective problem 
as a constrained optimization problem which maximizes classification accuracy subject to a
compression ratio constraint. We incorporate trainable 
JPEG compression blocks and JPEG decoding blocks together with the trainable CNN classifier in our 
end-to-end learning model. Our proposed DL 
model can configure 
JPEG encoding parameters to achieve
high classification accuracy.

We organize the rest of the paper as follows. 
Section \ref{chap:relatedwork} introduces the 
basics of JPEG codec and summarizes related works.
Section \ref{chap:framework} 
proposes the novel end-to-end DL architecture.
We provide experimental results in Section \ref{chap:experiments},
before concluding and discussing potential future directions in Section \ref{chap:conclusion}.

\section[JPEG Codec and Learning over JPEG]{JPEG Codec and Learning over JPEG}
\label{chap:relatedwork}

\subsection{JPEG Codec in View of Deep Learning}

In JPEG compression with 4:2:0 chroma subsampling, 
an RGB source image is first converted to YC{\scriptsize{B}}C{\scriptsize{R}} color space through linear transformations.  Chrominance channels (C{\scriptsize{B}} and C{\scriptsize{R}}) are subsampled by 2 both vertically and horizontally. 
After subsampling, each of the 3 YC{\scriptsize{B}}C{\scriptsize{R}} channels is split into non-overlapping $ 8\times 8 $ blocks before applying  blockwise DCT. 
The 2-dimensional (2-D) DCT of an image block $\boldsymbol{I}$ of size $N\times N$ with entries $\boldsymbol{I}(k,l)$ is defined by $N\times N$ block $\boldsymbol{F}= \boldsymbol{DID}^{T}$, where $ \boldsymbol{D} $ is a constant matrix.
The DCT is capable of compacting image features with 
a small number of DCT coefficients with little perceptible loss 
after compression.

For compression, each block of the $8\times 8$ frequency-domain coefficients $\boldsymbol{F}$ is quantized using pre-defined quantization matrices, or ``Q-tables" $\boldsymbol{Q}$ with entries $\boldsymbol{Q}(j,k)$ at JPEG
encoder to obtain quantized block $ \boldsymbol{F}_q $ whose entries are
$
\boldsymbol{F}_q(j,k) = \mbox{round}[{\boldsymbol{F}(j,k)}/{\boldsymbol{Q}(j,k)}]
$.
The decoder reconstructs from the compressed block $\boldsymbol{F}_q$ and the Q-table to form Hadamard product $\boldsymbol{F}_q=
\boldsymbol{F}\circ\boldsymbol{Q}$. Decoder would then use inverse DCT (IDCT) to recover spatial RGB images.
Parameters in 
$\boldsymbol{Q}$ can be adjusted to achieve different 
compression levels and visual effects. JPEG standard provides two Q-tables 
to adjust compression loss, one for the Y channel
and another for C{\scriptsize{B}} and C{\scriptsize{R}} channels.

In networked image applications, training using full resolution images would make little practical sense
and would be prone to accuracy loss because
only codec-compressed image data
are available at the cloud/edge processing node. Thus, 
deep learning networks should directly use compressed 
DCT coefficients as inputs for both
training and inference instead of full resolution
RGB images for training. Image classification (labeling) directly based on DCT coefficients can 
further reduce decoder computation
during both training and inference by 
skipping the IDCT and potentially achieve
better robustness under dynamic levels 
of JPEG compression.

Importantly, ResNets that were successfully developed for recognition of fully reconstructed JPEG images tend to exhibit performance loss if they
are directly used on image data in DCT domain.
Motivated by the need to improve image processing performance
in networked environments under channel bandwidth and storage constraints,
this work investigates deep learning architecture designs suitable for 
optimizing standard compliant JPEG configurations to achieve high
classification accuracy and low bandwidth consumption by
directly applying DCT input data. 
Our joint optimization of the JPEG configuration
is achieved by optimizing 
both the JPEG Q-table parameters and the deep learning classifier
to achieve end-to-end deep learning 
framework spanning from the IoT source encoder to the
cloud classifier. 
Our experiments include tests on the high resolution ImageNet dataset. 

\vspace*{-2mm}
\subsection{Related Works}
\vspace*{-1mm}
For bandwidth and storage conservation, DL architectures such as auto-encoders have been effectively trained\cite{singh2020end,1703.00395,1701.06487}
for image compression with little degradation of
classification accuracy or perceptual quality. Previous works \cite{7026196,NIPS2018_7649} also revealed
direct training of DL models on the DCT coefficients 
using faster CNN structure with a modified ResNet-50 architecture. 
The authors of \cite{quannet}
developed a joint compression and classification network model based
on JPEG2000 encoding. These works suggest the benefit of 
DL-based end-to-end optimization of image codecs.

Previous studies \cite{Liu_2018,li2020optimizing,choi2020task,luo2020rate} have also recognized the importance of Q-tables in JPEG codecs and seek to optimize them for DL-based image classification. \cite{li2020optimizing,Liu_2018} propose to design JPEG Q-tables based on the importance of DCT coefficients, evaluated by the relative
frequency \cite{li2020optimizing} or the standard deviation \cite{Liu_2018} of the coefficients. 
Both \cite{choi2020task,luo2020rate} offer end-to-end 
DL models to
estimate a set of optimized Q-tables for each 
input image, where models are pre-trained to predict the bandwidth of each image. 
In contrast, targeting low-cost sensing nodes, our proposed model 
learns a single set of Q-tables for all 
images during training, which can be pre-configured within the JPEG codec after training
for
inference tasks. This reduces the required
computational power at the sensing nodes.
Moreover,
our proposed training tunes Q-tables
and does not require a separate entropy estimation model.

To our best knowledge, there exists no published work on JPEG Q-table optimization
for distributed learning to target low-cost sensing devices. Since JPEG continues to be a commonly used image coding methods in massive number of low-cost devices, 
we focus our investigation on 
the rate-accuracy trade-off to facilitate their
widespread applications in distributed learning. 

\vspace*{-2mm}
\section{Joint DL Architectures} \label{chap:framework}
\vspace*{-1mm}

\subsection{Wide ResNet (WRN) for CIFAR-100 and Tiny ImageNet}
\begin{figure*}[tbhp]
	\centering
	\includegraphics[width=0.9\linewidth]{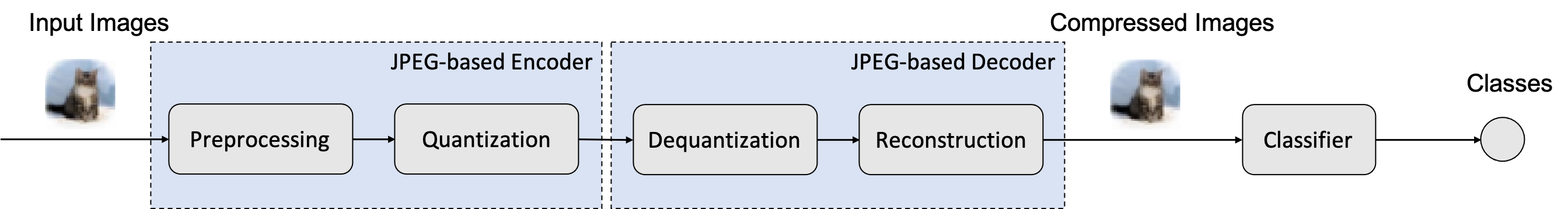}
	\caption{Proposed architecture for CIFAR-100 and Tiny ImageNet. The detailed structure in WRN-28 is omitted here.}
	\label{fig:cifar_wrn}
		\vspace*{-3mm}
\end{figure*}

\vspace*{-1mm}
For CIFAR-100 and Tiny ImageNet, we propose
the WRN model of Fig.~\ref{fig:cifar_wrn}. Following JPEG , our
preprocessing steps include level shifting, color transformation, subsampling, and DCT. The invertible color transformation can 
also be trained.

\label{chap:wrn}

\subsubsection{Compression Layers}
The trainable quantization layer in our proposed WRN model is similar to the ``quan block'' in \cite{quannet}, 
as shown in Fig.~\ref{fig:cifar_wrn}. JPEG can take three distinct Q-tables $\boldsymbol{Q}_1$, $\boldsymbol{Q}_2$, and $\boldsymbol{Q}_3$, respectively, for each of the YC{\scriptsize{B}}C{\scriptsize{R}} color channels, 
leading to 192 parameters in this layer. Same as \cite{quannet}, we replace the trainable quantization parameters by element-wise reciprocal of 
the Q-table entries $
\boldsymbol{q}_{i}(j,k) = \boldsymbol{Q}_{i}(j,k)^{-1}.$
The matrices $\boldsymbol{q}_{i}$, referred to as the ``compression kernels'', allows the model to learn to discard any frequency domain information by setting the corresponding entry $ \boldsymbol{q}_{i}(j,k) = 0 $. Smaller $ \boldsymbol{q}_{i} $ values lead to smaller range of quantized DCT coefficients and consequently generates
fewer encoded bits.

The quantization layer includes a non-differentiable rounding operation $a(\boldsymbol{F}) = \mbox{round}(\boldsymbol{F})$, which cannot be used in a gradient-based training framework, as its activation function. Following \cite{theis2017lossy}, we address this problem by substituting a smooth approximation  $\hat{a}(\boldsymbol{F}) = \boldsymbol{F}$ for the rounding function in  backpropagation.
Together, preprocessing and quantization layers form a JPEG-based encoder.

The dequantization layer only needs to multiply the 
encoded DCT blocks $\boldsymbol{F}_q$ element-wise by 
their respective Q-table matrices $\boldsymbol{Q}_{1}$, $\boldsymbol{Q}_{2}$
and $\boldsymbol{Q}_3$ 
corresponding to the encoder quantization layer.
The quantization and dequantization layers jointly
form a pair of ``compression layers''.

\subsubsection{Reconstruction Layer} 
As described in Section \ref{chap:relatedwork}, the reconstruction layer performs IDCT via $\boldsymbol{I} = \boldsymbol{D}^{T}\boldsymbol{F}_q\boldsymbol{D}$ for each quantized DCT coefficient block $\boldsymbol{F}_q$ and rearranges the reconstructed blocks $\boldsymbol{I}$. C{\scriptsize{B}} and C{\scriptsize{R}} 
channels are upsampled via bilinear interpolation. The outputs of this layer are  YC{\scriptsize{B}}C{\scriptsize{R}} spatial images. Together, the dequantization layer and the reconstruction layer form the JPEG decoder.

\subsubsection{Classifier}
WRNs\cite{zagoruyko2016wide} achieve impressive classification performance on the CIFAR-100\cite{Krizhevsky09learningmultiple} and Tiny ImageNet\cite{le2015tiny} datasets. Without loss of generality,
we adopt a 28-layer WRN as the classifier. For CIFAR-100, we set the convolutional layer width multiplier $k = 10$, same as that used in \cite{zagoruyko2016wide}. For Tiny ImageNet, we set $k = 1$ to further simplify training.

\subsection{Loss Function During Training}
To jointly reconfigure the
JPEG parameters in compression 
layers and to optimize the deep learning classifier,
we design the following loss function during
training:\vspace*{-2mm}
\begin{align*}
\mathcal{L} = \mathcal{L}_{\rm CLA} + \lambda \mathcal{L}_{\rm Quan},
\vspace*{-2mm}
\end{align*}
where $\mathcal{L}_{\rm CLA}$ is the cross entropy classification loss, $\mathcal{L}_{\rm Quan}$ is a penalty term for the compression kernels and $\lambda$ is an adjustable hyper-parameter to 
govern the importance of the rate. Since there is no simple or standard metric to directly control JPEG 
encoded image size, which further involves
RLE and Huffman coding, we propose the following surrogate 
penalty function:
\begin{align*}
        \mathcal{L}_{\rm Quan} = \sum_{i=1}^{3}\sum_{j=1}^{8}\sum_{k=1}^{8}& \left[\mbox{max}(\boldsymbol{q}_{i}^{2}(j,k)-c,0)
        + \lambda_1 |\boldsymbol{q}_{i}(j,k)| \right]
\end{align*}
where $c$ and $\lambda_1$ are tunable 
hyper-parameters. The $\ell_1$ loss term promotes 
sparsity whereas the $\ell_2$ loss term regulates
the compression kernels. The hyper-parameter $c$ 
acts as a constraint on the squared magnitude of $\boldsymbol{q}_{i}$, shall be appropriately selected based on the values of $\lambda$ and $\lambda_1$. Larger $\lambda$ and $\lambda_1$ and a smaller $c$ leads to higher compression ratio and lower classification accuracy.  We propose the 
current form of the surrogate penalty function 
after testing both logarithm and sigmoid functions
without witnessing performance benefits. 

\subsection{Modified ResNet for ImageNet}
\begin{figure*}[htbp]
	\centering
	\includegraphics[width=0.92\linewidth]{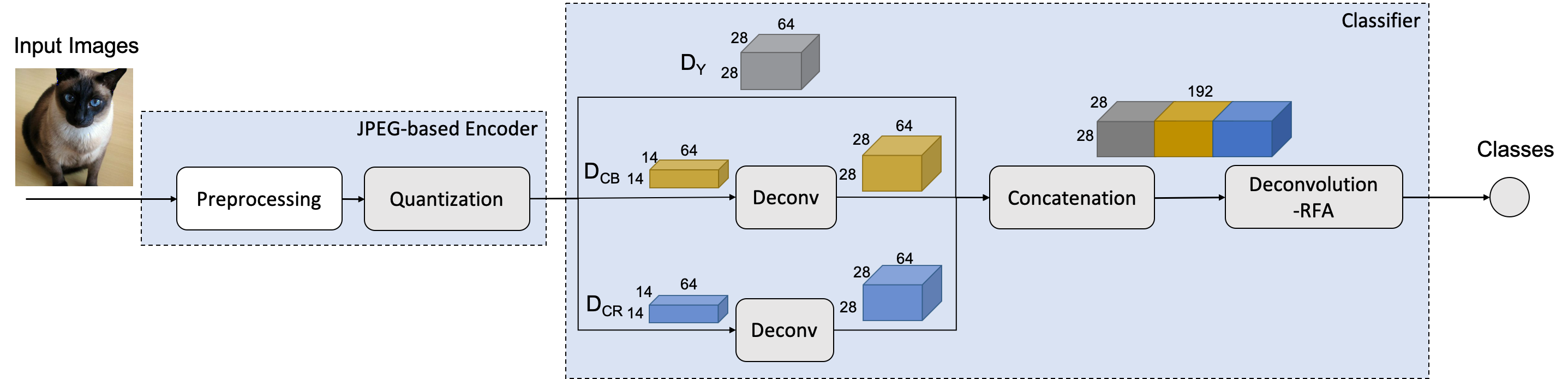}
	\caption{Architecture of our proposed model for ImageNet. The detailed structure of Deconvolution-RFA\cite{NIPS2018_7649} is omitted here.}
	\label{fig:ImageNet_resnet}
		\vspace*{-4mm}
\end{figure*}

For ImageNet, we adopt the same preprocessing steps and quantization layer from \ref{chap:wrn} and utilize the Deconvolution-RFA architecture in \cite{NIPS2018_7649}, which is inspired by ResNet-50, as the classifier. As demonstrated in Fig.~\ref{fig:ImageNet_resnet}, the quantized DCT coefficients of C{\scriptsize{B}} and C{\scriptsize{R}} channels are augmented to the same spatial size as Y channel by two separate 
transposed convolutional layers. The three channels 
are concatenated as input of the deconvolution-RFA model.
Considering the higher complexity of this model, 
we suggest a single $\ell_2$ regularization for compression kernels
for optimizing the quantization parameters via quantization loss $
\mathcal{L}_{\rm Quan} = \sum_{i=1}^{3}\sum_{j=1}^{8}\sum_{k=1}^{8}\boldsymbol{q}_{i}^{2}(j,k).$

\subsection{Implementation} 
We test the learning framework in Fig.~\ref{fig:cifar_wrn} with CIFAR-100 and Tiny ImageNet datasets.
The images in CIFAR-100 dataset are of $ 32\times 32 $ pixels, while images in Tiny ImageNet are of $ 64\times 64 $ pixels.

There are 192 trainable 
parameters in the compression kernels, all initialized to 1. The randomly-initialized WRN-28 classifier is trained jointly with the parameters in compression kernels, as well as the color transformation coefficients if needed, using Adam optimizer with a batch size of 100. The training of WRN-28 proceeds alternatively: the classifier is trained for 2 epochs while JPEG-based layers 
are frozen, followed by the compression layer being trained for 1 epoch while freezing the classifier. The training takes 150 such alternations. The learning rate starts from 0.05, and is scaled by 0.1 and 0.01, at alternation 50 and 100, respectively.

We implement the larger learning framework in Fig.~\ref{fig:ImageNet_resnet} with ImageNet in which the images are of $ 224\times 224 $ pixels. 
Similarly, color transformation coefficients are initialized to JPEG standard and 
all 192 quantization parameters are initialized to 1. The color transformation, compression kernels, and
Deconvolution-RFA classifier are trained
from end to end by using
Adam optimizer with a batch size of 32. 
The learning rate of compression kernels starts from $1\times 10^{-8}$ while that for other parameters starts from 0.001. 
Both learning rates are scaled by 0.1, 0.01, and 0.001, at epoch 30, 60 and 80, respectively. 
The training takes 90 epochs.

For the quantized DCT coefficients using
the new Q-tables, we 
customize Huffman tables for bandwidth saving
by randomly choosing 50k images from the corresponding training set
to generate Huffman tables. The bandwidth of the
validation data set is measured by
combining the principles of RLE and Huffman coding.

\vspace*{-1mm}
\section{Experiments}
\vspace*{-1mm}
\label{chap:experiments}
Our experiments are conducted on 
Keras and TensorFlow. We utilize the two metrics to evaluate perceptual quality of images: peak signal-to-noise ratio (PSNR) and structural similarity index measure (SSIM) index.
We test the proposed joint compression and classification (JCC) frameworks on
three datasets: (a) CIFAR-100 dataset based on 50k training
images and 10k test images belonging to 100 categories; (b) Tiny ImageNet dataset based on 100k training images and 10k validation images belonging to 200 categories; 
(c) ImageNet dataset based on approximately 1.3M training images and 50k test images belonging to 1000 categories. 
\vspace*{-1mm}
\subsection{JPEG Standard Baseline} 
\vspace*{-1mm}
We first present baseline results
of images using standard JPEG algorithm
with 4:2:0 chroma subsampling. We initialize 
compression kernels $\boldsymbol{q}_{i}$, $ i = 1,\ 2,\ 3 $ 
using the Q-tables given in JPEG standard. 
In this baseline scenario,
we only adapt the classifier parameters
during training.

For CIFAR-100, we consider 9 different JPEG image qualities between $ 12.5$\% and $100\% $. For Tiny ImageNet, we select 4 different image qualities between $ 10$\% and $80\% $. For ImageNet, our experiments consider 5 different image qualities between $ 12.5$\% and $100\% $. 
The classification results are shown in
Figs.~\ref{fig:cifar100_results}-\ref{fig:ImageNet_results}. 
These baseline results reveal 
that the classification accuracy correlates positively  
with image qualities and average image bandwidth (rate).

\vspace*{-2mm}
\subsection{Joint Compression and Classification (JCC)}
\begin{figure*}[hbtp]
	\centering
	\subfloat{\includegraphics[width=0.33\linewidth]{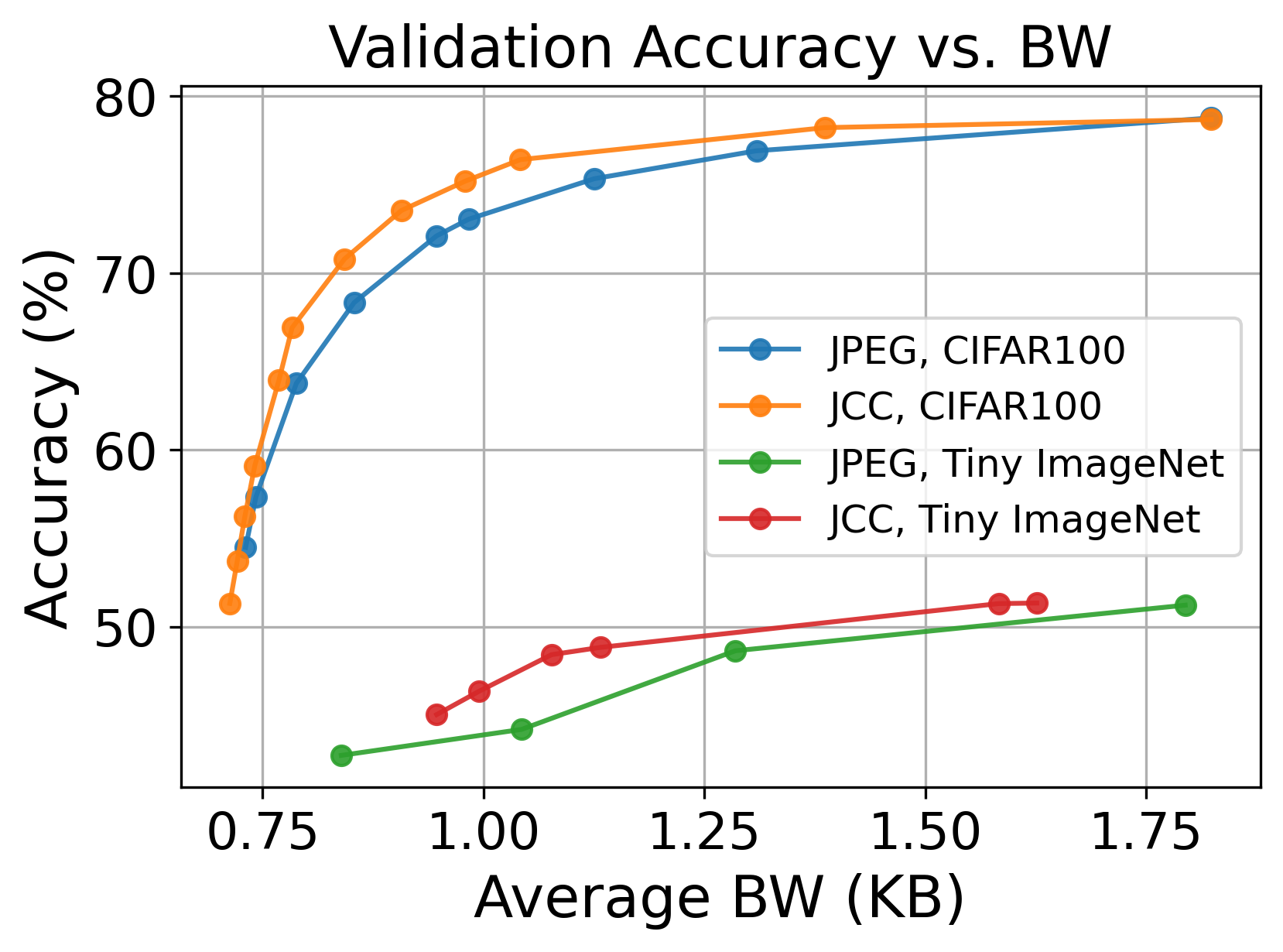}}
	\subfloat{\includegraphics[width=0.33\linewidth]{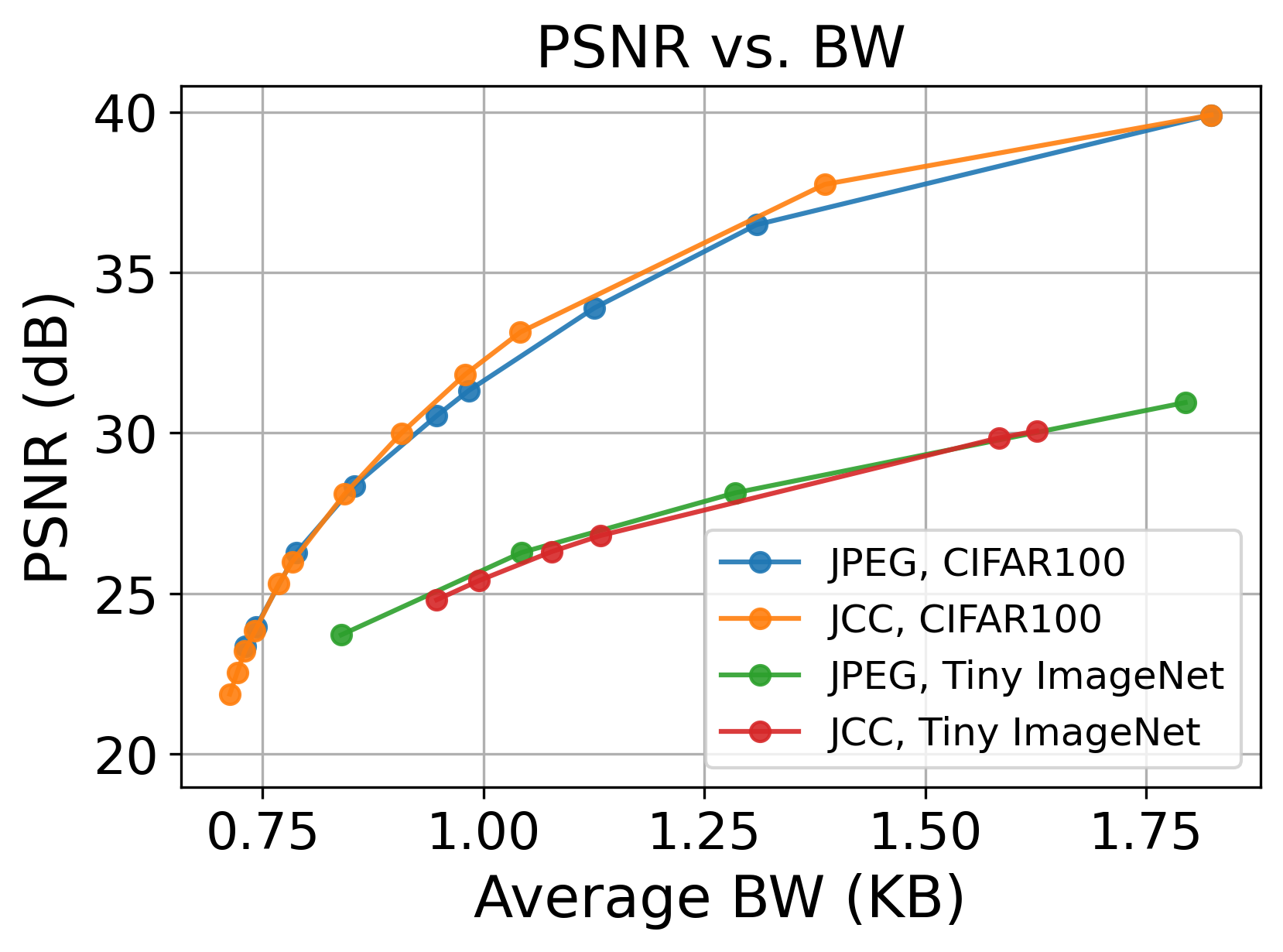}}
	\subfloat{\includegraphics[width=0.33\linewidth]{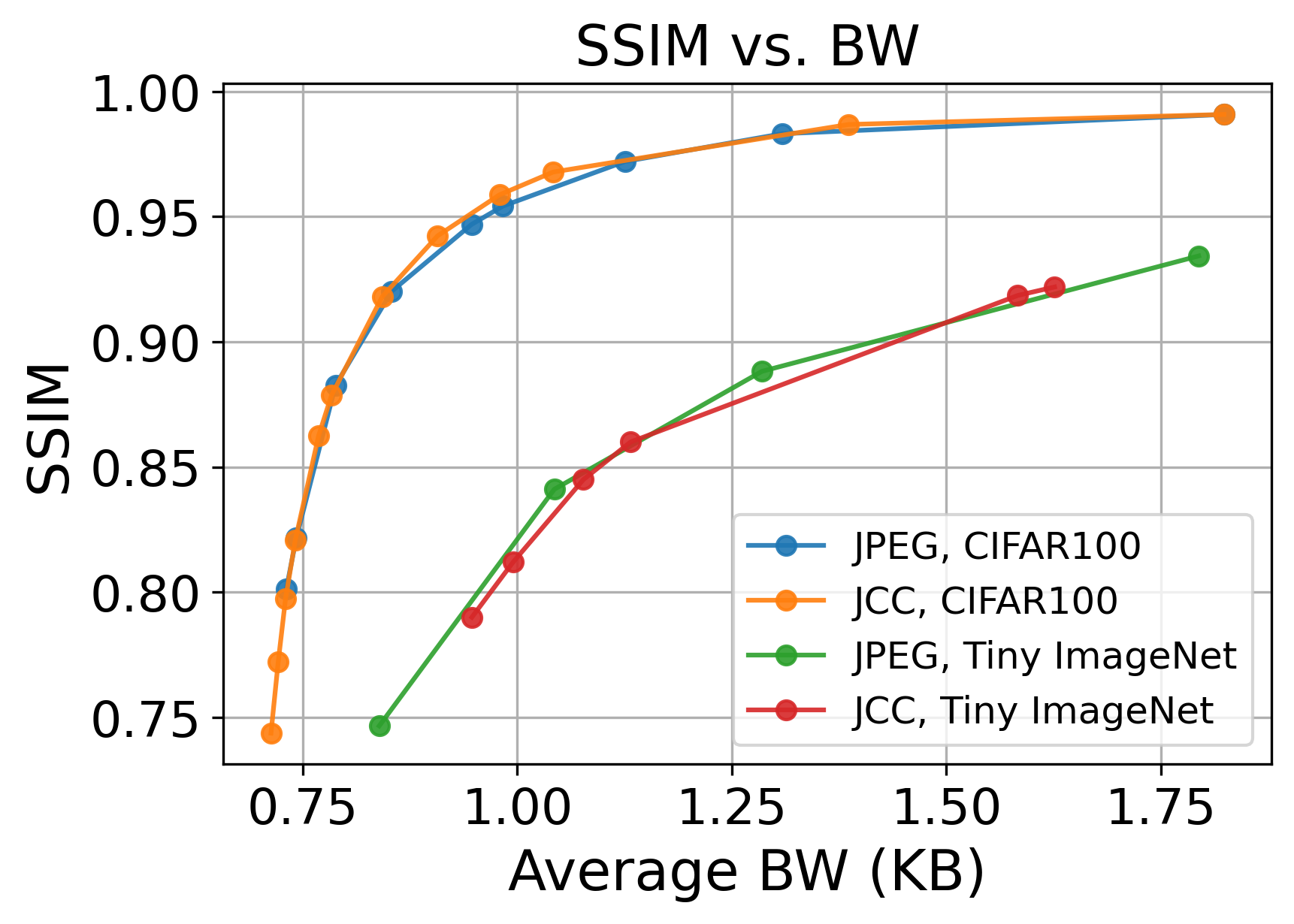}}
	\vspace*{-2mm}
	\caption{Comparison between JCC and JPEG standard on CIFAR-100 and Tiny ImageNet datasets with WRN-28 Model, with respect to classification accuracy, PSNR and SSIM versus average bandwidth.}
	\label{fig:cifar100_results}
		\vspace*{-4mm}
\end{figure*}

\vspace{-4pt}
For JCC, we train and optimize the compression kernels and the classifier. For CIFAR-100, we selected 8 values of $ \lambda $ from $ 10^{-6} $ to $5$ with $ \lambda_1 = 1$ and $c = 0.01/\lambda$. The classification accuracy, PSNR and SSIM results can be found in Fig.~\ref{fig:cifar100_results}. Compared with
the JPEG baseline, JCC achieves clear improvement of up to 
2.4\% in accuracy at bandwidth between 0.75 and 1.5 KB per image. The PSNR 
and SSIM of JCC-compressed images are similar to those using JPEG. 

For Tiny ImageNet, we select 6 values of $ c $ from $ 5\times 10^{-3} $ to $0.8$. As shown in Fig.~\ref{fig:cifar100_results}, when comparing with the JPEG baseline, we observe accuracy gain of up 
to 4\% by the proposed JCC model at low
bandwidth between 0.9 and 1.6 KB per image while maintaining similar visual quality. Overall, the PSNR and SSIM of JCC-optimized and JPEG standard quantization tables are similar.

\begin{figure*}[hbtp]
	\centering
	\subfloat{\includegraphics[width=0.33\linewidth]{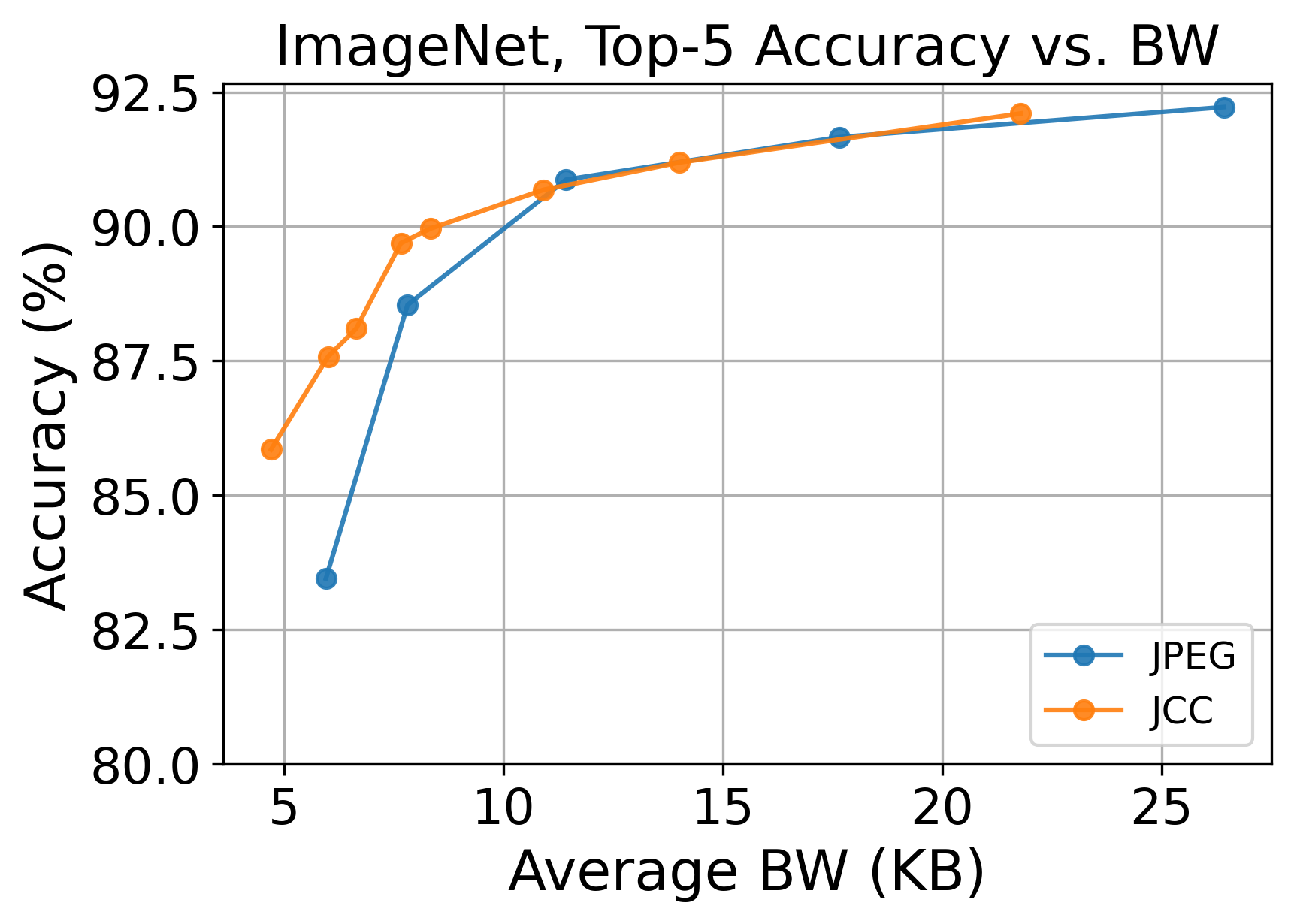}}
	\subfloat{\includegraphics[width=0.33\linewidth]{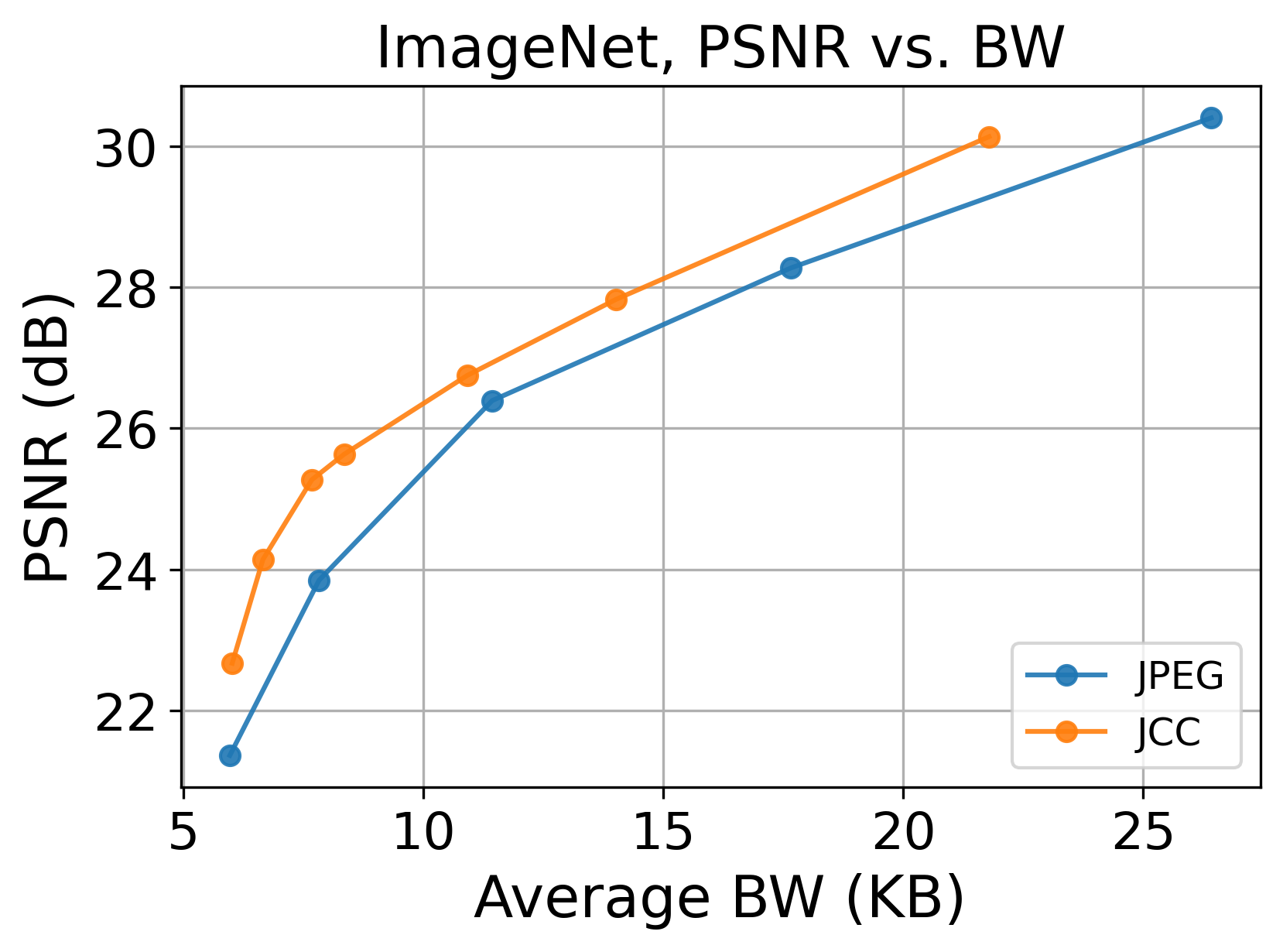}}
	\subfloat{\includegraphics[width=0.33\linewidth]{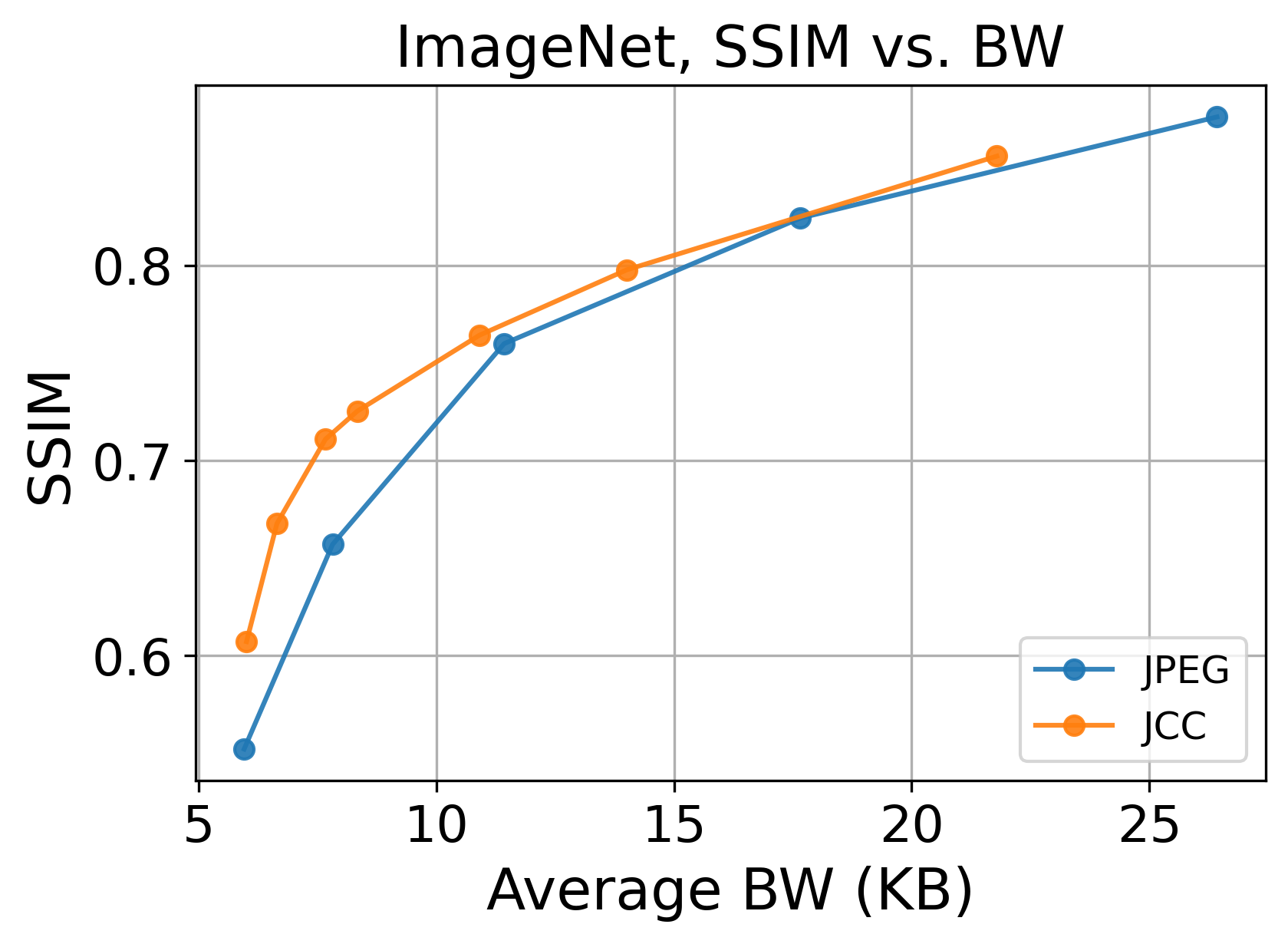}}
	\vspace*{-2mm}
	\caption{Comparison between JCC and JPEG baseline on ImageNet dataset with Deconvolution-RFA Model, with respect to top-5 classification accuracy, PSNR and SSIM versus average bandwidth. }
	\label{fig:ImageNet_results}
	\vspace*{-4mm}
\end{figure*}

For ImageNet, we consider 9 values of $\lambda$ 
between $25$ and $100$. 
The resulting top-5 classification accuracy, 
PSNR, and SSIM are given
in Fig.~\ref{fig:ImageNet_results}. For encoding
rates below 11 KB per image, the JCC model outperforms 
the baseline by up to $3.7\%$ in terms of classification accuracy. For bandwidths above 11 KB per image, 
the classification accuracy difference 
between JPEG and JCC is quite insignificant. 
Furthermore, PSNR and SSIM of JCC-compressed images outperform those of standard JPEG  encoded images.

From these experimental results, we observe that the JCC model can effectively optimize the JPEG compression kernels for better rate-accuracy trade-off, especially at 
moderate image bit rates.
It is intuitive that
the performance edge of JCC diminishes for 
very high image sizes because most image features can be preserved 
when given sufficient number of bits and JCC and JPEG no longer need to delicately balance the rate-accuracy trade-off. 
Similarly, for very low image sizes, very few bits can be used to encode vital information in DCT coefficients. Hence, the encoders have 
less flexibility to further optimize the
rate-accuracy trade-off, thereby making it difficult
for even the JCC model to find better parameter settings. 
\vspace*{-1mm}
\subsection{JCC and Color Transformation} \vspace*{-1mm}

Considering JCC and color transformation (JCC-color),
color transformation coefficients, compression kernels and classifier are jointly trained. We initialize the color conversion coefficients according to the
settings in JPEG. According to our experimental results, tuning the color transformation coefficients
offers no performance gain over JCC.
Theoretically, invertible color space transformation does not lead to information loss and can be subsumed by 
first dense-layer in the neural network. 
In fact, we observe that the
resulting color transform rarely move from
their initial values and further optimization is unnecessary.
\vspace*{-1mm}
\subsection{Further Analysis and Discussions}
\vspace*{-1mm}
\begin{figure}[htb]
    \centering
	\subfloat{\includegraphics[width=\linewidth]{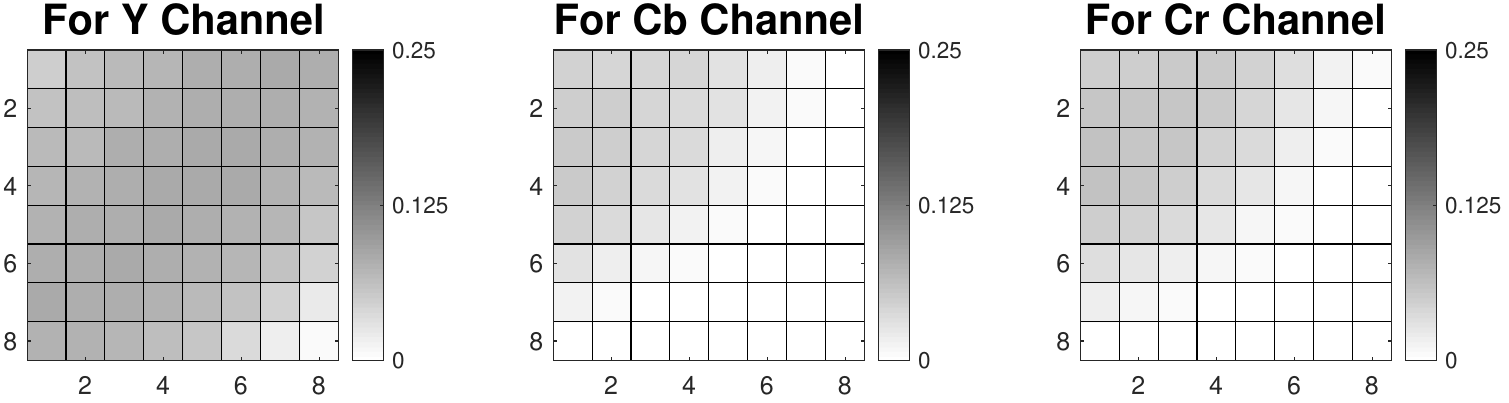}%
		\label{Quan_tb:train}}
	\caption{JCC optimized compression kernels based on CIFAR-100 for Y (left), C{\scriptsize{B}} (middle) and C{\scriptsize{R}} (right) channels, with $\lambda=0.001, \lambda_{1}=1$ and $c=10$.}
	\label{fig:Quan_tb}
	\vspace*{-6mm}
\end{figure}

Our experimental
results suggest that trainable JCC model can 
extract critical DCT features for classification among categories of CIFAR-100, 
Tiny ImageNet, and ImageNet datasets. Furthermore, the perceptual quality of images are preserved. Fig.~\ref{fig:Quan_tb} shows 
the resulting compression kernels  
with which WRN-28 achieves 
classification accuracy of 
75.20\% at an average rate of 0.979 KB/image. 
Darker grids imply low compression or higher importance of the corresponding DCT coefficient. 
The encoder clearly favors lower frequency bands.
Since there are longer consecutive 0's in the zig-zag order in the end-to-end learned compression kernels, the correspondingly compressed
DCT coefficients require fewer bits via RLE.  Furthermore, 
the trainable model learns to discard some higher frequency DCT components 
less critical to classification.

Together, these experimental results show performance enhancement
on standardized JPEG codec for cloud-based
image classification. The optimized Q tables
can be distributed in pre-installed JPEG encoders of low-cost devices 
through software updates for different encoding sizes. 
\vspace*{-1mm}
\section{Conclusions} \label{chap:conclusion}\vspace*{-1mm}
We present an end-to-end deep learning (DL)
architecture to jointly optimize JPEG
image compression and classification for low-cost
sensors in distributed
learning systems. Results on CIFAR-100, Tiny ImageNet, and ImageNet datasets 
demonstrate successful
training of the end-to-end DL framework 
for better image compression and classification performance without
perceptual quality loss. Optimized JPEG Q-tables can be 
readily incorporated within deployed codecs in practice. 
Future works may explore the broad appeal of this end-to-end learning principle in other bandwidth-constrained distributed 
DL tasks such as object detection,
segmentation, and 
tracking.  

\bibliographystyle{IEEEtran}
\bibliography{all_citations}

\end{document}